\documentclass{iopart}

\usepackage{iopams}
\usepackage{amsthm}
\usepackage{graphicx}

\newtheorem{Lemma}{Lemma}

\newtheorem{Corollary}{Corollary}

\begin{document}

%
% TITLE 
%
\title{Directed quantum communication}

%
% AUTHORS ETC.
%
\author{J {\AA}berg$^{1,2}$, S Hengl$^2$ and R Renner$^2$}
\address{$^1$ Institute for Physics, University of Freiburg, Hermann-Herder-Strasse 3, D-79104 Freiburg, Germany}
\address{$^2$ Institute for Theoretical Physics, ETH Zurich, 8093 Zurich, Switzerland}
\ead{johan.aberg@physik.uni-freiburg.de}

%%%%%%%%%%%%%%%%%%%%%%%%%%%%%%%%%%%%%%%%%
%		 ABSTRACT AND TITLE			             %
%%%%%%%%%%%%%%%%%%%%%%%%%%%%%%%%%%%%%%%%%

\begin{abstract}
We raise the question whether there is a way to characterize the quantum information transport properties of a medium or material. 
For this analysis the special features of quantum information have to be taken into account.  We find that quantum communication over an isotropic medium, as opposed to classical information transfer, requires the transmitter to direct the signal towards the receiver. Furthermore, for large classes of media there is a threshold, in the sense that `sufficiently much' of the signal has to be collected. Therefore, the medium's capacity for quantum communication can be characterized in terms of how the size of the transmitter and receiver has to scale with the transmission distance to maintain quantum information transmission. To demonstrate the applicability of this concept, an $n$-dimensional spin lattice is considered, yielding a sufficient scaling of $\delta^{n/3}$ with the distance $\delta$.
 \end{abstract}


\maketitle

%%%%%%%%%%%%%%%%%%%%%%%%%%%%%%%%%%%%%%%%%%
%			INTRODUCTION			   %
%%%%%%%%%%%%%%%%%%%%%%%%%%%%%%%%%%%%%%%%%%

\section{Introduction} 
The propagation of disturbances in materials, e.g., electric pulses in a piece of metal, sound in a solid, or spin-waves in a spin lattice, can be regarded as a transmission of information. Evidently, the `quality' of this information transmission is determined by the  transport properties of the medium. In this work we take an information-theoretic approach to transport properties, or perhaps more accurately, we regard the capacity for information transfer as a material property.

To get an intuitive picture of the setting we consider one can think of radio transmission over free space, i.e., imagine a propagation medium that is translation symmetric and isotropic (in a wide sense) and that we are in control only of limited transmitter and receiver regions. While radio transmission is typically modeled as classical information transfer over a classical medium, we here consider quantum information transfer over quantum mechanical media. Apart from the practical relevance of characterizing quantum information transfer properties for the purpose of quantum communication or processing in physical media, it is a fundamental theoretical issue to pinpoint how the special properties of quantum information alter the typical scenarios we know from classical communication theory. 

Here, we show that quantum communication in an isotropic medium, as opposed to classical information transfer, requires the transmitter to direct the signal towards the receiver, as one intuitively would expect from the no-cloning theorem. The degree to which such a directed quantum communication can be achieved is a property of the medium. We suggest to characterize this quantum information transport property by how the size of the transmitter and receiver regions have to scale with increasing transmission distance in order to obtain quantum communication. To the best of our knowledge, such characterizations have not been considered previously.

As an illustration we use an n-dimensional spin lattice, where an upper bound to the scaling can be determined. In the specific setting of spin lattices of higher dimensions (larger than $1$) this investigation can  be regarded as a generalization of the idea to use permanently coupled 1D spin chains for
information transmission \cite{Bose03,Bose07}. 
For 1D spin chains it is known that perfect state transfer can be
obtained by tuning the interactions locally along the chain \cite{Tuning}. One could imagine
this to be possible also in higher dimensions \cite{Casaccino2009}. However, as we consider the `free space' of a translation symmetric lattice, this excludes such local tunings.
In \cite{HeinTanner09} it was shown that communication between arbitrary points can be achieved without the transmitter and receiver knowing each others positions. However, this result assumes a finite lattice, which is excluded in our case by the effectively infinite medium. We also note that the propagation of information in a medium, as studied here, is related to the Lieb-Robinson bound \cite{Lieb1972}.\footnote{The Lieb-Robinson (LR) bound can be rephrased as an upper bound on the speed of information propagation. Reasonably, the LR bound should limit how efficiently quantum information can be transmitted in a medium.}

\section{Scaling characterization of media} 
\begin{figure}[t] 
\centering
\includegraphics{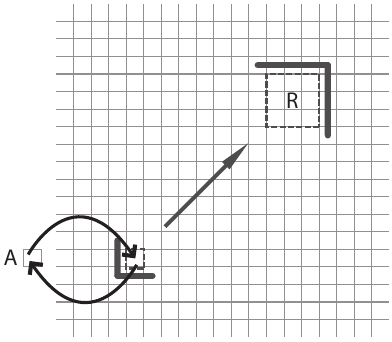} 
 \caption{\label{fig1:Lattice} 
To obtain a channel from a transmitter to a receiver over a spin lattice, we may use a separate spin $A$ as an input system. To `inject' this information into the lattice, we swap the input spin A with a selected spin in the lattice.
 A local potential barrier acts as a  transmitter antenna that directs the excitation towards the receiver,  where the wave packet reaches another antenna that collects the excitation into the receiver area. By considering the state in the receiver region $R$ at a given time we obtain a channel from the input spin $A$ to the receiver $R$. 
 }
\end{figure}

Quantum information transport is possible when the medium admits a non-zero quantum channel capacity. 
The latter measures how many qubits that can be sent reliably, when averaged over many independently repeated uses of a channel, assuming optimal encodings and decodings. (We consider the unassisted capacity, where, e.g., no additional classical channels are assumed.)
 To apply this concept we need to specify a channel, i.e., a well defined physical mapping from an input system to an output system. A channel can be set up by  `injecting' information from an input system $A$ into a bounded region of the medium, in the following referred to as the `transmitter region'. (For a concrete example in the special case of a spin lattice, see figure \ref{fig1:Lattice}.) If the input system $A$ initially is uncorrelated with the medium, then the injection and the evolution of the medium result in a quantum channel from $A$ to a receiver region $R$. 
One could imagine a qualitative characterization of the medium simply by asking whether the resulting channel capacity is non-zero or not. However, the answer will depend on the sizes and distance between the transmitter and receiver.
To avoid this, we rather ask how the transmitter and receiver regions have to \emph{scale} with the transmission distance to obtain a non-zero capacity. (To use scaling as a method to get rid of unimportant details is a common approach, e.g., in the context of area law scaling of entanglement entropy \cite{GroundStateEntang}.) The transmission still depends on other aspects of the information injection 
(and the extraction at the receiver) but the optimal scaling achievable (possibly under some constraints, e.g., a bound on the energy) can be taken as a characterization of the medium. 
Needless to say, the optimal scaling would in general be very challenging to determine. More realistically, we can find upper bounds (sufficient scaling) to the theoretically optimal scaling. (This is analogous to the classical setting where one in general has to settle for lower bounds on the channel capacity over a given medium.) With the purpose to obtain such scalings, we first elucidate some necessary and sufficient conditions for a non-zero channel capacity. We begin with a simple argument which shows that if there is too much symmetry in the system, then the quantum channel capacity is zero.

\section{Need for symmetry breaking}
Classical signals can be copied and transmitted in all directions, e.g., in radio broadcasting, where the copying is done by ramping up the amplitude in the transmitter antenna.  Since quantum information cannot be cloned \cite{Wootters1982} or broadcast \cite{Barnum1996} one might suspect that there is no quantum analogue of this. We can make this intuition more precise in terms of a symmetry argument. For this purpose we assume the medium to have some type of symmetry, and furthermore assume that the state of the medium after the injection is invariant under this symmetry, for all states of the input system $A$. (Since we typically imagine a localized  transmitter, the symmetries would be, e.g., rotations or reflections around this region.)
 The symmetry generates copies of the receiver region $R$. If such a copy $R'$ does not overlap with $R$, then they correspond to two distinct subsystems of the medium. 
By the assumed symmetries, $R$ and $R'$ will obtain the same state no matter the input $A$. Intuitively, the no-cloning theorem thus implies that there is no quantum information transmission from $A$ to $R$. More formally, since the state of $R$ can be reconstructed from $R'$, this implies that the channel from $A$ to $R$ is anti-degradable  \cite{Caruso}, which gives a zero quantum channel capacity \cite{Caruso,Holevo} (see  \ref{Symmetrybreaking} for more details).
We can thus conclude that the symmetry makes quantum communication impossible. This is in contrast to the classical case, where a similar symmetry condition may lower the efficiency, but would not prevent information transmission per se. 

The above arguments show that symmetry breaking is a necessary condition for quantum communication. However, in the following we show that for large classes of systems this is not enough; the quantum signal needs to be directed in a stronger sense. Loosely speaking, we need to gather `sufficiently much' of the signal to achieve quantum information transmission. We begin by demonstrating this threshold effect in a very simple system.

\section{Thresholds for quantum communication: An illustration} 
Consider a medium where information is transmitted via single excitations or particles. 
(We do not specify whether the medium is discrete or a continuum.) In this setting one can determine a simple necessary and sufficient condition for a non-zero quantum channel capacity. 
We assume that the medium preserves the total number of particles, i.e., its Hamiltonian commutes with the total number operator. We furthermore assume that the medium has a vacuum state $|\nu\rangle$ that can be written as a product state $|\nu\rangle =  |0_{R}\rangle|0_{R}^{c}\rangle$ of local zero-excitation states $|0_{R}\rangle$ and $|0_{R}^{c}\rangle$ in the receiver $R$ and its complement $R^c$, respectively. Moreover, the single-excitation sector is spanned by states of the form $|\chi_R\rangle|0^c_{R}\rangle$ and $|0_R\rangle|\chi^c_{R}\rangle$, where  $|\chi_{R}\rangle$ and $|\chi_{R}^{c}\rangle$ are single excitation states on $R$ and $R^c$, respectively.\footnote{One can relax these assumptions. The vacuum does not have to be a product state, and it is essentially enough if the single-excitation sector is spanned by states that can be generated from the vacuum (and removed again) coherently, via local operations.} 
   
The input $A$ is a single qubit, the medium starts in the vacuum state, and the injection can be described by a unitary operator $U_{I}$. If $A$ is in state $|0\rangle$ then the injection does nothing, i.e., $U_{I}|0\rangle|\nu\rangle = |0\rangle|\nu\rangle$, while it puts a single excitation state, $|\eta_{T}\rangle$, in the transmitter region if  $A$ is in $|1\rangle$, i.e., $U_{I}|1\rangle|\nu\rangle = |0\rangle|\eta_{T}\rangle$.  
 The dynamics of the lattice evolves $|\eta_{T}\rangle$ into a new single-excitation state  $|\psi_{p}\rangle  =  \sqrt{p}|\chi_{R}\rangle|0_{R}^{c}\rangle + \sqrt{1-p}|0_{R}\rangle|\chi_{R}^{c}\rangle$, where $p$ is the probability to find the excitation in the receiver region.
If the state of the input qubit $A$ is $\rho$, then the state of the receiver region $R$ can be written as
\begin{eqnarray*}
\Phi_{p}(\rho) = & \langle 0|\rho|0\rangle|0_{R}\rangle\langle 0_{R}|  +  p\langle1|\rho|1\rangle|\chi_{R}\rangle\langle \chi_{R}| \\
&+ \sqrt{p}\langle 1|\rho|0\rangle|\chi_{R}\rangle\langle 0_{R}|  + \sqrt{p}\langle 0|\rho|1\rangle|0_{R}\rangle\langle \chi_{R}|  \\
& + (1-p)\langle 1|\rho|1\rangle|0_{R}\rangle\langle 0_{R}| .
\end{eqnarray*}
 Effectively, $\Phi_{p}$ is a qubit amplitude damping channel, and for these it is known that the channel capacity  is non-zero if and only if $p> 1/2$ \cite{AmplitudeDamping}. (For another example see \ref{restricted}.)
 If combined with the previous symmetry argument, we see that it is not enough to break the symmetry in order to get a non-zero capacity, but that the receiver furthermore has to collect most of the amplitude of the particle.

\section{Thresholds in sufficiently noisy transmissions}
In general media, a disturbance can be an arbitrarily complicated combination of multi-excitations that may decay or disperse relative to some, possibly noisy, equilibrium distribution, e.g., a thermal state of the medium. Here we show that under wide conditions, the quantum transmission still shows  threshold effects, which can be regarded as a channel version of `entanglement sudden death' \cite{Yu04}. 

 As we bring the transmitter and receiver further apart (assuming otherwise fixed setups), the state in the receiver should reasonably become less and less distinguishable from the background. In the limit of infinite distance the resulting channel would thus be the replacement map $\Lambda_{\sigma}(\rho) = \sigma$, for all input states $\rho$, where $\sigma$ is the reduced density operator of the receiver resulting from the equilibrium state of the medium. For finite distances, the difference between the actual channel $\Phi$ and the limiting channel $\Lambda_{\sigma}$ (e.g., as measured by the the `diamond norm' $\Vert \Phi -\Lambda_{\sigma}\Vert_{\diamond}$ \cite{Kitaev97,Aharonov97}) can thus be taken as a measure of the extent to which the actions of the transmitter can be distinguished from the background. (This quantity generalizes the role of the pick-up probability in the example above.)
If $\sigma$ is mixed enough to have full rank, then there exists a  neighborhood of $\Lambda_{\sigma}$ where \emph{all} channels have zero quantum channel capacity. (See \ref{thresholds}.) 
 This tells us that even if $\Vert \Phi -\Lambda_{\sigma}\Vert_{\diamond}$  never becomes identically zero as we increase the separation between transmitter and receiver, the resulting quantum channel capacity will nevertheless be zero beyond some threshold distance. This threshold can be increased if we increase the sizes of the transmitter and receiver regions. Thus, it is possible to characterize the medium in terms of the scaling of the transmitter and receiver regions needed to maintain a non-zero quantum channel capacity with increasing distance (see \ref{thresholddistance}). This is in contrast to the case of classical information transfer (over classical or quantum channels) where we generically would expect a non-zero (albeit small) classical capacity for all distances, which makes a characterization in terms of a scaling for a non-zero capacity  meaningless (see \ref{classical}).
  
\section{Possibility of directed quantum communication} 
To illustrate the possibility of directed quantum communication, we take a square lattice $L$ of uniformly coupled spin-half particles that interact according to the Heisenberg XY-model
\begin{equation}
\label{Heisenberg}
  H=-\frac{1}{2}\sum_{\langle j,k\rangle} \left(\sigma^x_j
    \sigma^x_{k} + \sigma^y_{j} \sigma^y_{k}\right)
  + \sum_{j}\left(\sigma^z_j+\hat{1}_j\right), 
\end{equation}
where $\sigma_j$ denotes Pauli-matrices at position $j$, and $\langle j,k\rangle$  nearest neighbor pairings. In the 1D case (allowing for varying coupling constants) this is a common model for information transfer in spin chains (see e.g. \cite{Tuning}).
Since $[H,\sum_j \sigma_j^z]=0$, the total number of excitations  is conserved, and the ground state is a product state $|0\rangle\cdots|0\rangle$ where $0$ denotes spin down. 
 The simple dynamics of this model facilitates numerical calculation of the pick-up probability (and thus the channel capacity).
Due to computational limitations we only consider the 2D case.
 
In \cite{Muelken2005} it was observed that a single excitation can propagate along diagonals of the 2D square lattice XY-model in a remarkably confined manner (see figure 7 in \cite{Muelken2005}). However, the wave packet disperses more rapidly in other directions. In other words, the pick-up probability in the receiver region and hence the channel capacity depends on the direction of propagation, similar to other transport properties. 
In the present calculations we consider propagation along the favored diagonals.

One can imagine several different methods to direct the excitations towards the receiver. One  way is to construct local potential barriers, as depicted in figure \ref{fig1:Lattice}. These potentials are obtained by adding terms of the form $w_{j}\sigma_j^z$ to (\ref{Heisenberg}), where $w_j$ are real numbers.
We use this simple type of antennas for the calculation of the dashed line in figure \ref{fig:oneexcitation} (a), which gives the pick-up probability $p$ as a function of the time $t$ between the swap-in from $A$ and the time when we record the state in $R$. As figure \ref{fig:oneexcitation} (a) shows, $p$ reaches above the critical value $1/2$ for this specific arrangement. Another method to obtain  the necessary directionality (which numerical tests suggest is superior to the antenna construction) is to put a suitably shaped wave packet directly on the lattice. The solid line in figure \ref{fig:oneexcitation} (a) gives one example of this for a modulated Gaussian wave packet cropped to a small  transmitter region.

\section{Sufficient scaling: An example}
\begin{figure}[t]
   \centering
   \includegraphics{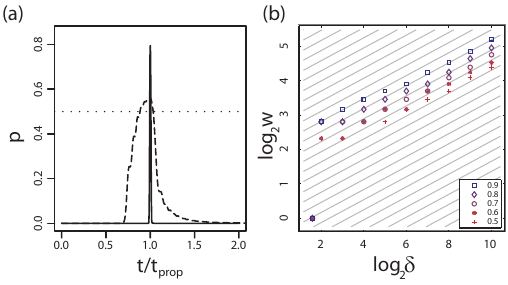}
     \caption{\label{fig:oneexcitation} 
  (a) Pick up probability: The probability $p$ to find the excitation in the receiver area is plotted as a function of the evolution time $t$ measured relative to the propagation time $t_{\textrm{prop}}$ of the peak. The dashed line corresponds to the setting schematically depicted in figure \ref{fig1:Lattice}, with a $256\times 256$ lattice with lossy edges,  and a $20\times 20$ receiver area. The distance between the inner corners of the antennas is 110 sites. The solid line corresponds to a $2048\times 2048$ lattice, with  $21\times 21$ transmitter and receiver regions. The distance between the centers of these two squares is 1969 sites. In this case we have no antennas, but use as initial state a suitably modulated Gaussian wave packet, cropped to the transmitter region. As seen, both cases reach above the critical value $1/2$. \\
 (b) Scaling: With a transmitter and receiver at distance of $\delta$ sites in the lattice, we let the initial wave package be a Gaussian, modulated to travel at the maximal group velocity, and cropped to a square transmitter region with a side length that scales as $\delta^{1/3}$. For this transmission system  we determine the side length $w$ of a square-shaped receiver region needed to obtain a given pick-up probability $p$, as a function of $\delta$. We plot  $\log_{2}w$  against $\log_{2}\delta$, and repeat this for the pick-up probabilities $p= 0.9, 0.8, 0.7, 0.6, 0.5$. The lines in the background are set to the slope $1/3$.
  }
\end{figure}

Using the above model, with transmission along the diagonals of the lattice, we here turn to the question of how fast the transmitter and receiver have to grow with the transmission distance to obtain a non-zero channel capacity.
A crucial issue is how fast a given single-particle wave-packet spreads as it propagates, and thus minimally dispersive wave-packets should be useful. 
For the 1D XY-model it was found \cite{Osborne04,Haselgrove05}  that  a good choice of such wave-packets yields a  pick-up probability close to $1$, for transmitter and receiver regions that grow like $\delta^{1/3}$, where $\delta$ is the number of spins in the spin chain.\footnote{For single excitations, the Heisenberg model in \cite{Osborne04,Haselgrove05} is equivalent to the Heisenberg XY-model we use.} This suggests an analogous approach for the XY-model on an $n$-dimensional square lattice, since the evolution is decoupled along the $n$ different dimensions, which would yield a volume scaling of $\delta^{n/3}$ of the transmitter and receiver regions. This reasoning is confirmed in figure \ref{fig:oneexcitation} (b) by a numerical calculation of the scaling in the 2D case. Since we have used a specific transmission system, this is an upper bound to the theoretically optimal scaling. However, restricted to the set of single-excitations, it appears reasonable to expect this result to be near optimal.

\section{Conclusions} We have found that quantum communication requires us to direct and collect sufficiently much of the signals into the receiver. 
This makes it possible to characterize the quantum information transport in media in terms of the scaling of the transmitter and receiver region needed to maintain a non-zero quantum channel capacity.
For single-particle transmission in an $n$-dimensional Heisenberg XY-model, a scaling of $\delta^{n/3}$ is sufficient.

It is an open question how the scalings of general physical media, e.g., solid state systems or optical lattices, depends on various aspects of the dynamics, especially if we incorporate more realistic settings and include, e.g., Anderson localization, thermal noise, and decoherence. To directly determine the optimal scalings appears challenging, but estimates for sufficient scalings appear tractable.

In this investigation we have made the tacit assumption that a sequence of transmissions can be described as independent and identically distributed (iid) repetitions of a single transmission. If the medium in some sense relaxes to its initial state after each transmission, this approximation is justifiable, as the scaling does not take into account the time it takes to transmit signals, thus allowing sufficient delays between subsequent transmissions.
However, if we wish to determine the transmission per time unit, rather than per channel use, the iid assumption may not be useful, e.g., as the number of excitations in the medium potentially increases for rapidly repeated transmissions. 
Techniques that go beyond the iid assumption  \cite{Kretschmann05,Bayat08,Giovannetti09,Datta09} could potentially be applied in this case.

%%%%%%%%%%%%%%%%%%%%%%%%%%%%%%%%%%%
%		ACKNOWLEDGEMENT	%
%%%%%%%%%%%%%%%%%%%%%%%%%%%%%%%%%%%

\ack

We acknowledge support from the Swiss National Science Foundation (SNF), grant nos.~200021-119868
and 200020-135048, and from the European Research Council (ERC), grant
no.~258932. J{\AA} also acknowledges support from the Excellence Initiative of
the German Federal and State Governments (grant ZUK 43).

%%%%%%%%%%%%%%%%%%%%%%%%%%%%%%%%%%%
%		APPENDIX			%
%%%%%%%%%%%%%%%%%%%%%%%%%%%%%%%%%%%

\appendix 

\section{\label{Symmetrybreaking}Quantum communication requires symmetry breaking}

In the main text we sketched in mere words the argument for why symmetry breaking is needed to obtain a non-zero quantum channel capacity. Here we make the argument a bit more precise, and we begin by recapitulating the notion of degradable and anti-degradable channels.  

Given a finite-dimensional Hilbert space $\mathcal{H}$, we let $\mathcal{L}(\mathcal{H})$ denote the set of linear operators on $\mathcal{H}$. We let $\mathrm{TPCPM}(\mathcal{H}_{i},\mathcal{H}_{f})$ denote the set of trace preserving completely positive maps (channels) from $\mathcal{L}(\mathcal{H}_{i})$ to $\mathcal{L}(\mathcal{H}_{f})$.

 For every channel $\Phi$ there exists a Steinspring dilation, i.e., a Hilbert-space $\mathcal{H}_{C}$ and a partial isometry $V:\mathcal{H}_{A}\rightarrow \mathcal{H}_B\otimes \mathcal{H}_{C}$,  with $V^{\dagger}V = \hat{1}_{A}$, such that 
 \begin{equation}
\Phi(\rho) = \tr_{C}(V\rho V^{\dagger}), \quad \forall \rho \in\mathcal{L}(\mathcal{H}_A).
 \end{equation}  
We define the complementary channel $\Phi^c$ to $\Phi$ as
\begin{equation}
\Phi^{c}(\rho) := \tr_{B}(V\rho V^{\dagger}), \quad \forall \rho \in\mathcal{L}(\mathcal{H}_A).
\end{equation} 
The original channel, $\Phi$, is called degradable if there exists some channel $\mathcal{N}:\mathcal{L}(\mathcal{H}_C)\rightarrow\mathcal{L}(\mathcal{H}_B)$ such that $\Phi^c = \mathcal{N}\circ\Phi$.
In other words, a channel is degradable if it is possible to reconstruct the state of  the output $C$ from the state of  $B$. Vice versa, $\Phi$ is called anti-degradable if there exists a channel $\Lambda$ such  that $\Phi = \Lambda\circ\Phi^c$.
It is a well know fact that anti-degradable channels have zero quantum channel capacity \cite{Caruso,Holevo}.

As described in the main text, we obtain a channel by `injecting' information from a system $A$, initially uncorrelated to the medium. For this we use a channel $\Phi_{i}^{T}:\mathcal{L}(\mathcal{H}_{T}\otimes \mathcal{H}_A)\rightarrow \mathcal{L}(\mathcal{H}_{T})$, where $T$ is the transmission region in the medium. (It is useful to include the transmitter region $T$ at the input of this channel, as this makes it possible  to handle cases where $T$  initially is correlated with an environment, or other parts of the medium.) 

After the propagation in the medium one can furthermore imagine to `eject' the information from the receiver into an output system $B$, by using a channel $\Phi_{e}^{R}:\mathcal{L}(\mathcal{H}_{R})\rightarrow \mathcal{L}(\mathcal{H}_{B})$. In most discussions we will simply use the partial trace $\tr_{R^c}$, i.e., we  consider the receiver region itself as the output system, although in some cases it can be convenient to use a separate output system and other maps. 

To model the propagation in the medium in full generality, we assume the medium $M$ to initially be in some joint state $\sigma_{ME}$ with an `environment' $E$. We furthermore assume the propagation to be described by a unitary $V_{ME}$ (as generated by some joint Hamiltonian $H_{ME}$). In total we can thus describe the resulting channel from the input to a receiver region $R$ as 
\begin{equation}
\label{totalmap}
\Phi^{R}(\rho_A) = \tr_{ER^{c}}[V_{EM} [\Phi_{i}^{T}\otimes I_{ET^c}](\sigma_{EM}\otimes \rho_{A}) V_{ME}^{\dagger}].
\end{equation}
Note that by including the environment $E$ we allow all types of decay, decoherence, and noise effects.

In the following we wish to express the idea that the medium and the injection possess a symmetry. Let us therefore consider some symmetry group  $G$ with a unitary representation $\{U_g\}_{g\in G}$ on the Hilbert space  $\mathcal{H}_{ME}$ of the medium. (A reasonable special case is to let $U_{g} = \tilde{U}_g\otimes \hat{1}_E$, with $\tilde{U}_{g}$ only acting on $\mathcal{H}_{M}$.) What we need  is that the state of the medium, after the injection and the evolution, is invariant under the action of the group, irrespective of what state we feed to the input system. In other words, 
\begin{eqnarray}
\label{nldmvna} & \tr_{E}\big[U_{g}V_{EM}  [\Phi_{i}^{T}\otimes I_{ET^c}](\sigma_{EM}\otimes \rho_{A}) V_{ME}^{\dagger}U_{g}^{\dagger}\big]\\
\nonumber & \quad= \tr_{E}\big[V_{EM} [\Phi_{i}^{T}\otimes I_{ET^c}](\sigma_{EM}\otimes \rho_{A}) V_{ME}^{\dagger}\big],
\end{eqnarray}
for all $g\in G$ and all $\rho\in\mathcal{L}(\mathcal{H}_A)$. 
We discuss this assumption further below, but for the moment, let us assume that (\ref{nldmvna}) holds. 

Given a subsystem $R$ in the medium, every element $g$ of the group maps $R$ to a new subsystem $gR$. Assume that $R$ and $gR$ are two independent subsystems, i.e., we can decompose the total Hilbert space of the medium as $\mathcal{H}_{M} = \mathcal{H}_{R}\otimes \mathcal{H}_{gR}\otimes\mathcal{H}_{\textrm{leftovers}}$. In our setting, $R$ is a bounded region in the medium and $g$ is a symmetry operation, like a reflection or a rotation, why the necessary independence is obtained when the regions $R$ and $gR$ have no overlap. Considering the maps $\Phi^{R}$ and $\Phi^{gR}$ as in (\ref{totalmap}), for independent subsystems $R$ and $gR$, and assuming the symmetry condition (\ref{nldmvna}) to be true, it follows directly that $\Phi^{R}$ and $\Phi^{gR}$ are isomorphic. Consequently, both of them are anti-degradable and thus have zero quantum channel capacity. 

Although (\ref{nldmvna}) gives a clear condition, it might nevertheless good to illustrate it with a couple of extreme cases. In the simplest case we do not include any environment, and thus only consider unitary evolution generated by a Hamiltonian $H$ of the medium.  In this case $U_g$ of course only acts on $\mathcal{H}_{M}$ and the symmetry of the medium is guaranteed by $[U_g,H] = 0$ for all $g\in G$. That the information injection always results in a symmetric state, we can express as $U_g[\Phi_{i}^{T}\otimes I_{T^c}](\sigma_{M}\otimes \rho_A)U_g^{\dagger} = [\Phi_{i}^{T}\otimes I_{T^c}](\sigma_{M}\otimes \rho_A)$ for all $g$ and all $\rho_{A}$.
These two assumptions yield the `environment-free' special case of (\ref{nldmvna}).

Another extreme case is to assume that the environment is Markovian, e.g., replacing the Hamiltonian evolution by a Markovian master equation  on the medium alone \cite{Lindblad76,Kossakowski72}. 
The unitary operator $V_{ME}$ describing a time-step of the medium is thus replaced by a channel $\mathcal{E}:\mathcal{L}(\mathcal{H}_{M})\rightarrow \mathcal{L}(\mathcal{H}_M)$. The symmetry of the injection is expressed identically as in the previous example, while the symmetry of the evolution can be stated as $U_g\mathcal{E}(\rho)U_g^{\dagger} = \mathcal{E}(U_g\rho U_g^{\dagger})$. This yields a `channel  version' of (\ref{nldmvna}).

%%%%%%%%%%%%%%%%%%%%
%		Condition				%
%%%%%%%%%%%%%%%%%%%%

\section{\label{thresholds} Zero quantum capacity neighbourhoods around full rank replacement maps}

In the main text we claimed that for each full rank replacement map on finite-dimensional Hilbert spaces there exists a neighbourhood where all channels have zero quantum channel capacity. (Given a  replacement map $\Lambda_{\sigma}(\rho) = \sigma$ we say that $\Lambda_{\sigma}$ is `full rank' whenever $\sigma$ has full rank. We also say that $\Lambda_{\sigma}$ is `rank-deficient' if $\sigma$ is not full rank.)  As mentioned in the main text, this can be viewed a channel-analogue of what sometimes is referred to as `entanglement sudden death' (ESD) \cite{Yu04}. The rather extensive literature on this subject (see e.g. \cite{Coffman00, Zyczkowski01, Doisi03, Dodd04, Yu04, Santos06, Cunha07}) in essence  shows that the entanglement in many decoherence models can reach zero after a finite time.
The link to channel capacities is apparent, and one can translate results from ESD to the present setting using the Choi isomorphism \cite{Choi}. However, here we directly use the PPT criteria to obtain  a radius around full rank replacement maps, within which all channels have zero quantum channel capacity. This can be used for an upper bound to the scaling-characterization described in \ref{thresholddistance}. (In \ref{restricted} we also discuss thresholds in restricted neighbourhoods of rank-deficient replacement maps.)

We let $\mathrm{Lin}(\mathcal{H}_i,\mathcal{H}_f)$ denote the set of all linear maps from $\mathcal{L}(\mathcal{H}_i)$ to $\mathcal{L}(\mathcal{H}_f)$. 
Given an orthonormal basis $\{|j\rangle\}_{j=1}^{N}$ of $\mathcal{H}_{i}$ the Choi representation \cite{Choi} of an element $\Phi\in \mathrm{Lin}(\mathcal{H}_i,\mathcal{H}_f)$ is defined by, 
\begin{equation}
\mathrm{M}(\Phi) := \frac{1}{N}\sum_{j,j'}\Phi(|j\rangle\langle j'|)\otimes |j\rangle\langle j'|.
\end{equation}

\begin{Lemma}[\cite{Horodecki,Peres}]
\label{pptcondition} 
If the Choi-representation of a channel $\Phi$ has a positive partial transpose (we say that $\Phi$ is a PPT channel) then $\Phi$ has zero quantum channel capacity.
\end{Lemma}

For any linear operator $Q$ we denote the standard operator norm as $\Vert Q\Vert := \sup_{\Vert \psi\Vert = 1}\Vert Q|\psi\rangle\Vert$, the trace norm $\Vert Q\Vert_{1} := \tr\sqrt{Q^{\dagger}Q}$, and the Hilbert-Schmidt norm $\Vert Q\Vert_2 := \sqrt{\Tr(Q^{\dagger}Q)}$. For any $\Phi\in\mathrm{Lin}(\mathcal{H}_{i},\mathcal{H}_{f})$ we can define the diamond norm \cite{Kitaev97, Aharonov97}, as 
\begin{equation}
\Vert \Phi\Vert_{\diamond} := \sup_{X\in\mathcal{L}(\mathcal{H}_i\otimes\mathcal{H}_c):\Vert X\Vert_1\leq 1}\Vert [\Phi\otimes I_c](X) \Vert_{1},
\end{equation}
where $\dim(\mathcal{H}_c)\geq \dim(\mathcal{H}_i)$ \cite{Aharonov97}. 

Given an orthonormal basis $\{|k\rangle\}_{k}$ of the Hilbert space $\mathcal{H}$, we define the transpose of of an operator $Q$ on $\mathcal{H}$ as $\Theta(Q):= \sum_{k,k'}|k\rangle\langle k'|Q|k\rangle\langle k'|$.
We let $\mathrm{HP}(\mathcal{H}_{i},\mathcal{H}_{f})$ denote the set of Hermiticity preserving linear maps from $\mathcal{L}(\mathcal{H}_{i})$ to $\mathcal{L}(\mathcal{H}_{f})$. For any $\Phi\in \mathrm{HP}(\mathcal{H}_{i},\mathcal{H}_{f})$ we define
\begin{equation}
\xi(\Phi) := \lambda_{\mathrm{min}}\boldsymbol{(}\Theta_{f} \mathrm{M}(\Phi)\boldsymbol{)},
\end{equation}
where $\lambda_{\mathrm{min}}$ is the smallest eigenvalue of $\Theta_f \mathrm{M}(\Phi)$.
This is essentially the negativity \cite{Vidal02} of the state $M(\Phi)$.
Note that it does not matter whether we use $\Theta_{f}$ or $\Theta_{i}$ in the definition. 
Furthermore, if $\Theta_f \mathrm{M}(\Phi)$ has negative eigenvalues, then $\xi(\Phi)$ is to be understood as its most negative eigenvalue.

If $A$ is a Hermitian operator, we let $\lambda^{\downarrow}(A)$ denote the eigenvalues of $A$ in non-increasing order, i.e., $\lambda^{\downarrow}_{1}(A)Ê\geq \lambda^{\downarrow}_{2}(A)\geq \cdots \geq \lambda^{\downarrow}_{N}(A)$. 
\begin{Lemma}[Theorem VIII.4.8 in \cite{Bhatia}]
\label{ndsflkbn}
Let $A$ and $B$ be  Hermitian operators on the same finite-dimensional Hilbert space. Then
$\max_{j}|\lambda_{j}^{\downarrow}(A)-\lambda_{j}^{\downarrow}(B)| \leq \Vert A -B\Vert$.
\end{Lemma}

\begin{Lemma}
\label{NormsOfTranspose}
Regarded as a linear map, the partial transpose, $\Theta_f\otimes I_{i}$, satisfies the following properties:
\begin{eqnarray}
\label{eqv2norm} & \sup_{X\in\mathcal{L}(\mathcal{H}_{f}\otimes\mathcal{H}_{i}): \Vert X\Vert_{1}\leq 1}\Vert \Theta_f\otimes I_{i}(X)\Vert_{2} = 1,\\
\label{eqvinfnorm} & \sup_{X\in\mathcal{L}(\mathcal{H}_f\otimes\mathcal{H}_{i}): \Vert X\Vert_{1}\leq 1}\Vert \Theta_f\otimes I_{i}(X)\Vert \leq 1.
\end{eqnarray}
\end{Lemma}
The left hand side of (\ref{eqvinfnorm}) should not be confused with the completely bounded norm \cite{Paulsen} of $\Theta_f$, which would be obtained if we replaced the condition $\Vert X\Vert_{1}\leq 1$ with $\Vert X\Vert\leq 1$ (for $\dim\mathcal{H}_{i}\geq \dim\mathcal{H}_f$).

\begin{proof}
If $|\alpha\rangle,|\beta\rangle\in\mathcal{H}_{f}\otimes\mathcal{H}_{i}$ are normalized, one can use the Schmidt-decomposition to show that 
\begin{equation}
\label{nkblvns}
\Vert [\Theta\otimes I](|\alpha\rangle\langle \beta|) \Vert_{2} = 1.
\end{equation}
Let $X\in\mathcal{L}(\mathcal{H}_f\otimes\mathcal{H}_i)$ be such that $\Vert X\Vert_{1} \leq 1$.
Using a singular value decomposition $X = \sum_{n}s_{n}|\alpha_{n}\rangle\langle \beta_{n}|$, together with (\ref{nkblvns}), yield $\Vert [\Theta\otimes I](X) \Vert_{2}\leq  1$, which proves (\ref{eqv2norm}). Due to the general fact that $\Vert \cdot\Vert \leq \Vert \cdot\Vert_{2}$ we can conclude that (\ref{eqvinfnorm}) also holds.
\end{proof}

\begin{Lemma}
\label{continuity}
Let $\Phi,\Psi\in \mathrm{HP}(\mathcal{H}_{i},\mathcal{H}_{f})$ then
\begin{equation}
|\xi(\Phi)-\xi(\Psi)| \leq \Vert \Phi-\Lambda\Vert_{\diamond}
\end{equation}
\end{Lemma}
\begin{proof}
\begin{eqnarray*}
|\xi(\Phi)-\xi(\Psi)|  & \leq & \max_{j}\big|  \lambda_{j}^{\downarrow}\big(\Theta_{f} \mathrm{M}(\Phi)\big)-\lambda_{j}^{\downarrow}\big(\Theta_{f} \mathrm{M}(\Psi) \big) \big| \\
  & \leq & \Vert \Theta_{f} \mathrm{M}(\Phi) -\Theta_{f} \mathrm{M}(\Psi) \Vert \\
  & \leq &  \Vert  \mathrm{M}(\Phi - \Psi) \Vert_{1}\\
  & \leq & \Vert \Phi-\Psi\Vert_{\diamond}.
\end{eqnarray*}
The first inequality follows trivially from the definition of $\xi(\cdot)$, the second from Lemma \ref{ndsflkbn}, and the third follows from (\ref{eqvinfnorm}) in Lemma \ref{NormsOfTranspose}
\end{proof}

\begin{Corollary}
\label{PPTneighbourhood}
If $\Psi\in \mathrm{TPCPM}(\mathcal{H}_{i},\mathcal{H}_{f})$ is such that $\xi(\Psi)\geq 0$, then all channels with a distance in the diamond norm less than or equal to $\xi(\Psi)$ to $\Psi$ are PPT.
 
If $\xi(\Psi) < 0$, then all channels with a distance in the diamond norm strictly less than $-\xi(\Psi)$ to $\Psi$ are in the complement of PPT.
\end{Corollary}

\begin{Corollary}
Let $\Lambda_{\sigma}\in\mathrm{TPCPM}(\mathcal{H}_{i},\mathcal{H}_{f})$ be the replacement map $\Lambda_{\sigma}(\rho) = \sigma$. Then all elements in $\mathrm{TPCPM}(\mathcal{H}_{i},\mathcal{H}_{f})$ with a distance to $\Lambda_{\sigma}$ less than or equal to $\lambda_{\mathrm{min}}(\sigma)/\dim\mathcal{H}_i$, in the diamond norm, are PPT.
\end{Corollary}

\section{\label{restricted} Restricted zero quantum capacity neighbourhoods }

In the previous appendix we focussed on full rank replacement maps. It is certainly reasonable to ask if also rank-deficient replacement maps have zero quantum capacity neighbourhoods. We do unfortunately not provide an answer here, but merely observe that there exist physically relevant \emph{restricted} families of channels within which there still exist thresholds around low rank replacement maps.

One simple example is the family of channels $\Phi_p$, corresponding to the single excitation transmission described in the main text. In this case the relevant replacement map is $\Lambda_{|0\rangle\langle 0|}$, which is clearly not full rank. However, as was shown in the main text, within this (very restricted) class of channels, there is a threshold for zero capacity at $p = 1/2$. 

We can also obtain a multi-excitation generalization of this example. 
We let $|\eta_{T}\rangle$ be a $N$-excitation state, rather than a single-excitation state. Thus, after the evolution, the new $N$-particle state can be written  $|\psi\rangle = \sqrt{p}|0_{R}^c\rangle|\chi^{N}_{R}\rangle + \sqrt{q}|\chi^{N}_{R^c}\rangle|0_{R}\rangle +\sqrt{r}|\chi\rangle$, where $|\chi^{N}_{R}\rangle$ is an $N$-particle state in the receiver  region, $|\chi^{N}_{R^c}\rangle$ an $N$-particle state in the complement, and $|\chi\rangle$ is an $N$-particle state with more than zero excitations in both the receiver and the complement.  Here, $p$ is the probability that we find all the excitations in the receiver, $q$ the probability that we find none in the receiver, and $r = 1-p-q$ is the probability that we find some particles in both the receiver and its complement. The channel from the input qubit $A$ to the receiver $R$ can be written 
\begin{eqnarray*}
\Phi(\rho) = & \langle 0|\rho|0\rangle|0_{R}\rangle\langle 0_{R}|  +  p\langle1|\rho|1\rangle|\chi^{N}_{R}\rangle\langle \chi^{N}_{R}| \\
& + \sqrt{p}\langle 1|\rho|0\rangle|\chi^{N}_{R}\rangle\langle 0_{R}| + \sqrt{p}\langle 0|\rho|1\rangle|0_{R}\rangle\langle \chi^{N}_{R}| \\
& + q\langle 1|\rho|1\rangle|0_{R}\rangle\langle 0_{R}|  + r\langle 1|\rho|1\rangle \sigma,
\end{eqnarray*}
 where $\sigma$ is a density operator with support on the orthogonal complement to the space spanned by $|0_{R}\rangle$ and $|\chi^{N}_{R}\rangle$. Using degradability  and anti-degradability one can prove that this channel has a non-zero channel capacity if and only if $p>q$. Hence, the channel capacity is non-zero if and only if the probability to pick up all the particles is strictly larger than the probability to pick up none.  Like for the single-particle transmission we thus obtain a threshold effect for the channel capacity.

\section{\label{thresholddistance}Threshold distances and scaling}
Here we argue that the results of the previous appendices translate into the existence of a threshold distance for quantum communication. 

Imagine an effectively infinite, in some sense isotropic, and at least two-dimensional  medium, where  bounded transmitter and receiver are embedded.
 Assume that we fix the size and shape of the transmitter and receiver regions, as well as the injection and ejection maps, while we are allowed to vary the distance between the transmitter and receiver. It appears reasonable to assume that as we increase this distance, the transmitted signal will gradually fade away, and in the limit of an infinite distance, the receiver perceives only the background noise of the medium. Another way of putting this is to say that the actual transmission channel $\Phi$ approaches a replacement map $\Lambda_{\sigma}$. The state $\sigma$ is the image of the equilibrium state $\rho_{\textrm{eq}}$ of the medium in the receiver, i.e., $\sigma = \Phi^{R}_{e}(\rho_{\textrm{eq}})$, or simply $\sigma = \tr_{R^c}\rho_{\textrm{eq}}$.  Note that we do not necessarily require that the medium globally reaches an equilibrium state $\rho_{\textrm{eq}}$; it is enough if  the receiver sees something that locally looks like the equilibrium $\rho_{eq}$. One example of the latter is the single-excitation model in the main text, where $\rho_{\textrm{eq}} = |\nu\rangle\langle\nu|$. In this case, the excitation eventually will propagate away from the receiver, out into the effectively infinite medium. In other words, even though globally there can be an excitation present in the medium, it eventually will look to the receiver as if the medium is empty. 

Apart from the assumption that $\Phi$  approaches a limiting replacement map $\Lambda_{\sigma}$ as we increase the distance, we also assume that $\Lambda_{\sigma}$ is full rank. As shown in \ref{thresholds}, this implies that $\Phi$ for some sufficiently large distance eventually will enter the zero capacity neighbourhood indefinitely. 
Beyond this threshold distance the quantum channel capacity is identically zero. (As an alternative to the full-rank assumption, we may also use restricted models as in \ref{restricted}.) 

When we consider the question of how the size of the transmitter and receiver have to scale with increasing distance to maintain a non-zero capacity, it maybe goes without saying that we implicitly mean the scaling of regular and reasonably shaped regions. As we increase the size of the (e.g., sphere-shaped) transmitter or receiver regions, the threshold distance increases monotonically. This is due to the fact that we always can restrict ourselves to only use the original smaller region. One should also note that every medium has at least a trivial scaling. The reason is that we always can make the transmitter and receiver regions so large that they overlap. This would allow the transmitter to directly put the signal into the receiver, and thus trivially obtain perfect transmission. Hence, in the worst case, the radius of the transmitter or receiver scale linearly with the distance.

Given the above arguments it is reasonable to ask when we can expect  $\sigma$ to be a full rank operator. (Although one should keep in mind that it is not clear whether the full-rank assumption is essential, or merely an artifact of limited proof techniques.) For example, if $\rho_{\textrm{eq}}$ is the ground state of the Hamiltonian of the medium, then $\sigma = \tr_{R^c}\rho_{\textrm{eq}}$ is full rank when the ground state is sufficiently entangled between $R$ and $R^{c}$, in the sense of having the maximal Schmidt-rank. Another example is when $\rho_{\textrm{eq}}$ would be the Gibbs state of the medium Hamiltonian (i.e. $\rho_{\textrm{eq}} = e^{-\beta H}/Z(\beta)$). Whenever the underlying Hilbert space is finite-dimensional the Gibbs state has to be full rank.

\section{\label{classical}Generic non-zero classical capacity of quantum channels}
Here we briefly clarify the statement in the main text that the classical channel capacity in the neighbourhood of a replacement map generically is non-zero. This follows from the fact that a quantum channel has zero classical capacity if and only if it is a replacement map. 
A direct consequence of this is that every neighbourhood of a replacement map consists almost only of channels with non-zero classical capacity. The following proves that only replacement maps that have zero classical capacity.

The Holevo quantity \cite{Holevo98,Schumacher97} of a channel is defined as
\begin{eqnarray*}
&\chi(\Phi) :=  \sup_{p_x,\rho_x}\chi(\{p_x\}_x,\{\rho_x\}_x,\Phi),\\
&\chi(\{p_x\}_x,\{\rho_x\}_x,\Phi) :=  H\boldsymbol{(}\Phi(\sum_{x}p_x\rho_x)\boldsymbol{)}-\sum_xp_xH\boldsymbol{(}\Phi(\rho_x)\boldsymbol{)},
\end{eqnarray*}
 where the supremum is taken over all possible $p_x \geq 0$, $\sum_{x}p_{x} = 1$, and density operators $\rho_{x}$ in the domain of the channel.
 The classical capacity $C(\Phi)$ of a quantum channel $\Phi$ is the regularized version of the Holevo quantity  $C(\Phi) = \lim_{n\rightarrow\infty}\chi(\Phi^{\otimes n})/n$ \cite{Holevo98,Schumacher97}.
 Clearly, every replacement map has zero capacity. For the converse, we note that $C(\Phi) \geq \chi(\Phi)\geq 0$. Furthermore, 
$\chi(\{p_x\}_x,\{\rho_x\}_x,\Phi) = H(X:B)_{\widetilde{\rho}} \geq 0$, 
where $ H(X:B)_{\widetilde{\rho}} := H(\rho_{B})+H(\rho_X)-H(\rho_{XB})$ is the mutual information between $X$ and $B$ in the state $\widetilde{\rho} := \sum_{x}p_x|x\rangle\langle x|\otimes \Phi(\rho_{x})$,  where $\{|x\rangle\}_{x}$ is an orthonormal basis in an auxiliary Hilbert space $\mathcal{H}_{X}$. By the condition $\chi(\Phi) = 0$ we thus find that $H(X:B)_{\widetilde{\rho}}=0$ for all states $\widetilde{\rho}$. We can conclude that $\widetilde{\rho}$ must be a product state for all choices of $p_x$ and $\rho_x$, and thus $\Phi$ is a replacement map.

%%%%%%%%%%%%%%%%%%%%%%%%%%%%%%%%%%%
%           REFERENCES		  %
%%%%%%%%%%%%%%%%%%%%%%%%%%%%%%%%%%%

\end{document}